\documentclass[a4paper]{article}
\usepackage{graphicx}

\begin{document}

\begin{center}
{\bf \Large
Simulation of majority rule disturbed by power-law noise}
\bigskip

{\large
D. Stauffer*$^{\dag}$
and
K. Ku{\l}akowski$^{\ddag}$
}
\bigskip

{\em
Faculty of Physics and Applied Computer Science,
AGH University of Science and Technology,
al. Mickiewicza 30, PL-30059 Krak\'ow, Euroland\\
\medskip
* visiting from: Institute of Theoretical Physics, Cologne University,\\
D-50923 K\"oln, Euroland.
}

\bigskip
$^\dag${\tt stauffer@thp.uni-koeln.de}, 
$^\ddag${\tt kulakowski@novell.ftj.agh.edu.pl}

\bigskip
\today
\end{center}

\begin{abstract}
Simulations are reported on the Ising two-dimensional ferromagnet in the
presence of a special kind of noise. The noise spectrum $P(n)$ follows a
power law, where $P(n)$ is the probability of flipping randomly selected $n$
spins at each timestep. This is introduced to mimic the self-organized
criticality as a model influence of a complex environment.
We reproduced the phase transition similar to the case of $P(n)$ = constant.
Above some value of the noise amplitude the magnetisation tends to zero;
otherwise it remains constant after some relaxation. Information of the
initial spin orientation remains preserved to some extent by short-range
spin-spin correlations. The distribution of the times between flips is
exponential. The results are discussed as a step towards modeling of social
systems.

\end{abstract}



\bigskip

\section{Introduction}

Ising magnets and their variants have been simulated for opinion dynamics 
\cite{liggett,sznajd}, urban segregation \cite{schelling}, economics 
\cite{hohnisch}, language change \cite{nettle,wichmann} $\dots$, sometimes
even at zero temperature. In the latter case, each spin = $\pm 1$ aligns
in the direction of the majority of its neighbours; if the neighbourhood
is evenly divided we orient the spin randomly (Glauber kinetics). Starting
from a random distribution one does not always end up with all spins parallel;
strip domains may form at zero temperature \cite{spirin}. 

Noise can be introduced into this model also different from the traditional
temperature and Boltzmann probabilities. At each iteration of a $L \times L$ 
square 
lattice with four neighbours for each site, besides the above majority rule, 
we select $n$ times randomly a spin and flip it: ``noise'' \cite{oliveira,
jones}. The probability distribution function $P(n)$ of these numbers $n$ 
is taken as a power law, 
$$ P(n) \propto 1/n^{\alpha} \quad .\eqno (1) $$

To get this distribution, we determined random numbers $r$, homogeneously
distributed between zero and one, and then took
$$ n = T L^2 r^{1/(1-\alpha)}   \quad (1 \le n \le L^2)    \eqno(2a) $$
$$ n = T \exp(r \ln(L^2))            \eqno(2b) $$
for $\alpha$ smaller than  and equal to one, respectively. Here the 
proportionality factor $T$ determines the amplitude of the noise,
analogously to the temperature in Boltzmann statistics.

\begin{figure}
\begin{center}
\includegraphics[scale=0.5,angle=-90]{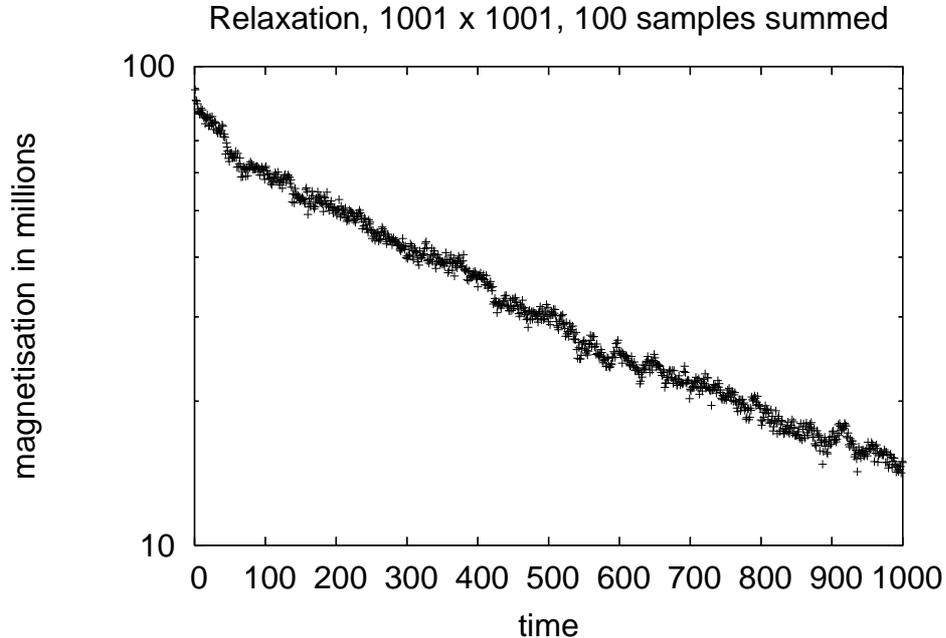}\\
\end{center}
\caption{Summed magnetisations versus time, $T=\alpha=1$.}
\end{figure}

\begin{figure}
\begin{center}
\includegraphics[scale=0.5,angle=-90]{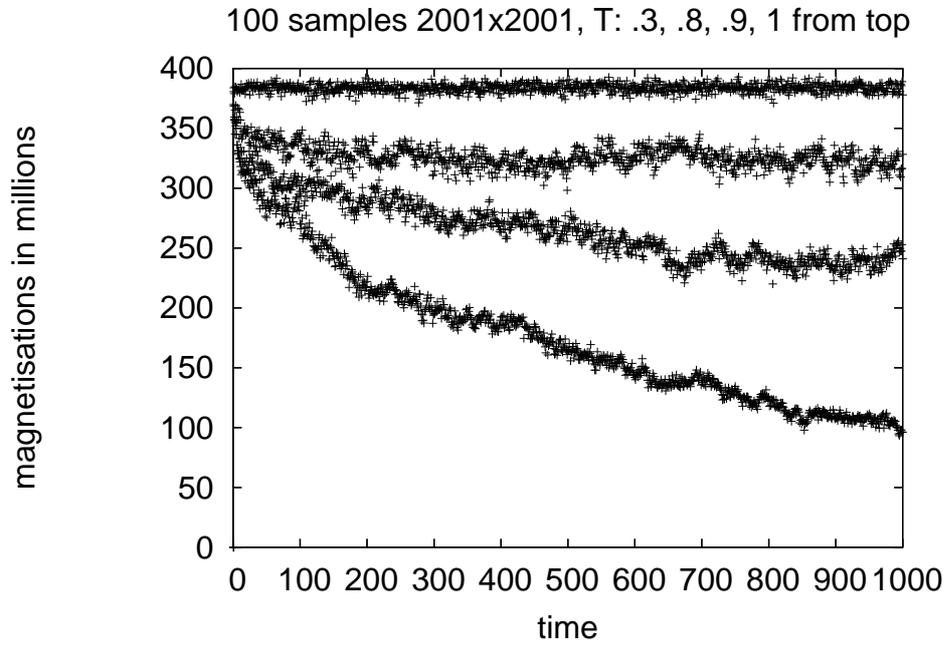}\\
\end{center}
\caption{Summed magnetisations versus time for $0.3 \le T \le 1,\; \alpha = 1$.}
\end{figure}

\begin{figure}
\begin{center}
\includegraphics[scale=0.45,angle=-90]{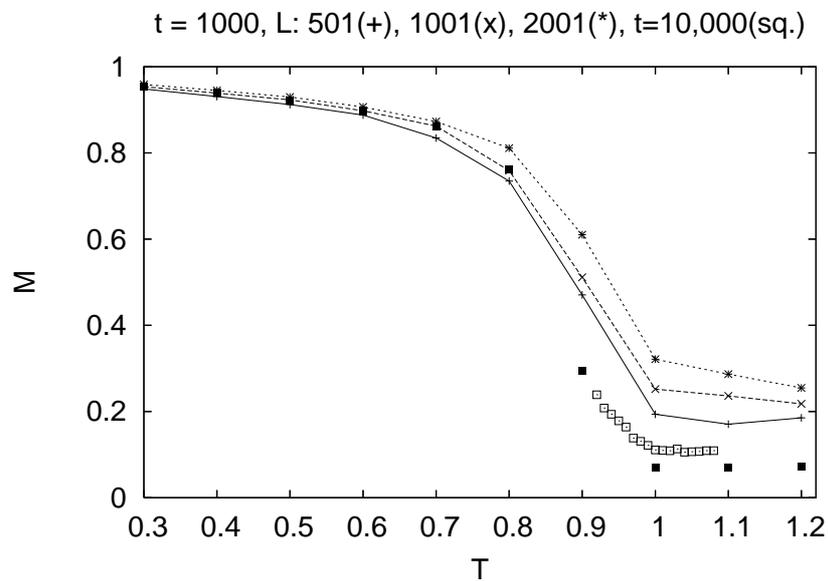}\\
\end{center}
\caption{Normalized magnetisations, averaged over 100 samples and over 
$500 < t \le 1000$. The open and full squares for $t = 10,000$  correspond
to $L = 501$ and 1001, respectively.}
\end{figure}

\begin{figure}
\begin{center}
\includegraphics[scale=0.45,angle=-90]{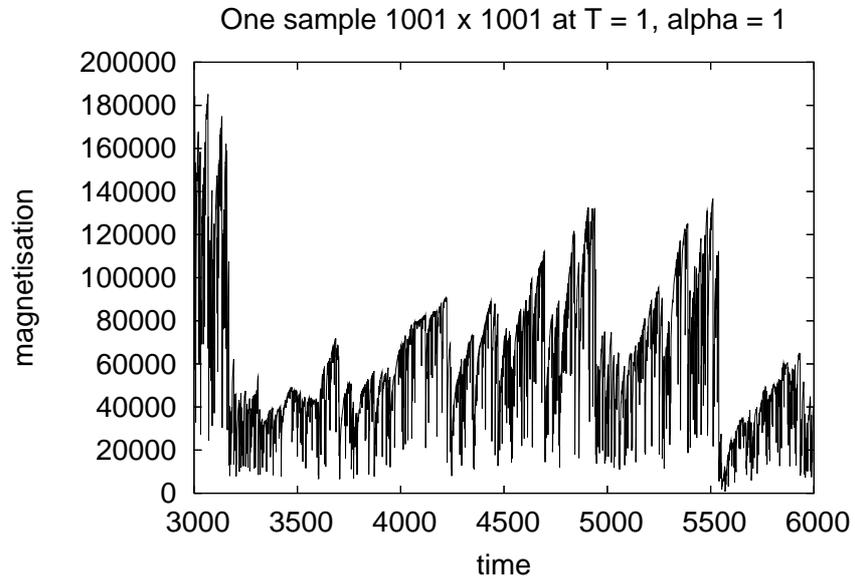}\\
\end{center}
\caption{Complicated time dependence is seen when looking at one sample only.}
\end{figure}

\begin{figure}
\begin{center}
\includegraphics[scale=0.45,angle=-90]{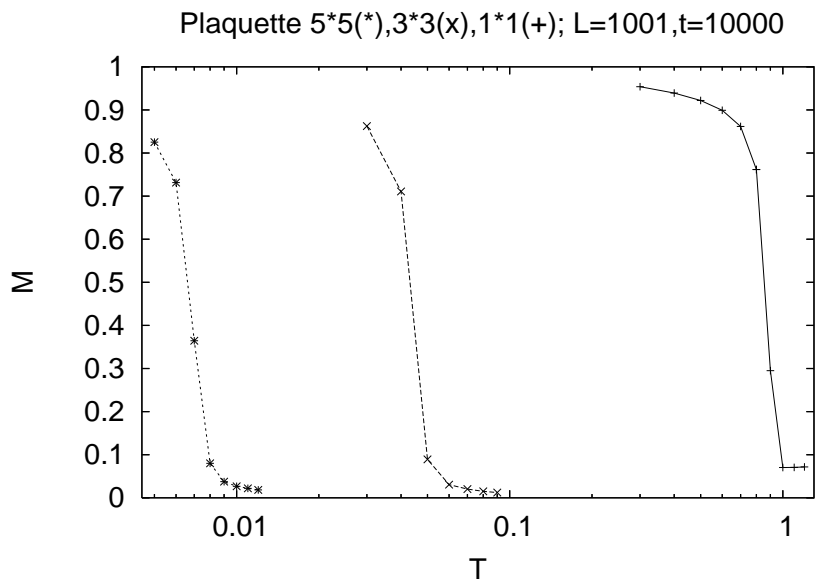}\\
\end{center}
\caption{Comparison of the standard case (+) with flipping plaquettes of
size $3 \times 3$ (x) and $5 \times 5$ (*). The $T$ axis is logarithmic.}
\end{figure}

\begin{figure}
\begin{center}
\includegraphics[scale=0.45,angle=-90]{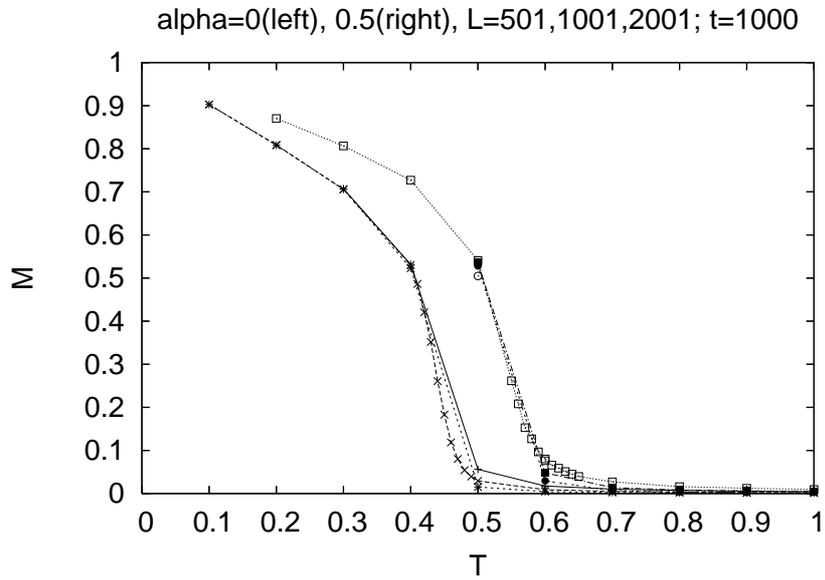}\\
\end{center}
\caption{Different $\alpha$, no plaquetttes: Normalized magnetisations, averaged
over 100 samples and over $500 < t \le 1000$. (The isolated circle refers 
to $5,000 < t < 10,000$.) Left curves $L = 501$ (+), 1001 (x), 2001 (*)
for $\alpha=0$ (case of \cite{oliveira}); right curves (open squares, full 
squares, full circles) same for $\alpha = 0.5$.
} 
\end{figure}

\begin{figure}
\begin{center}
\includegraphics[scale=0.45,angle=-90]{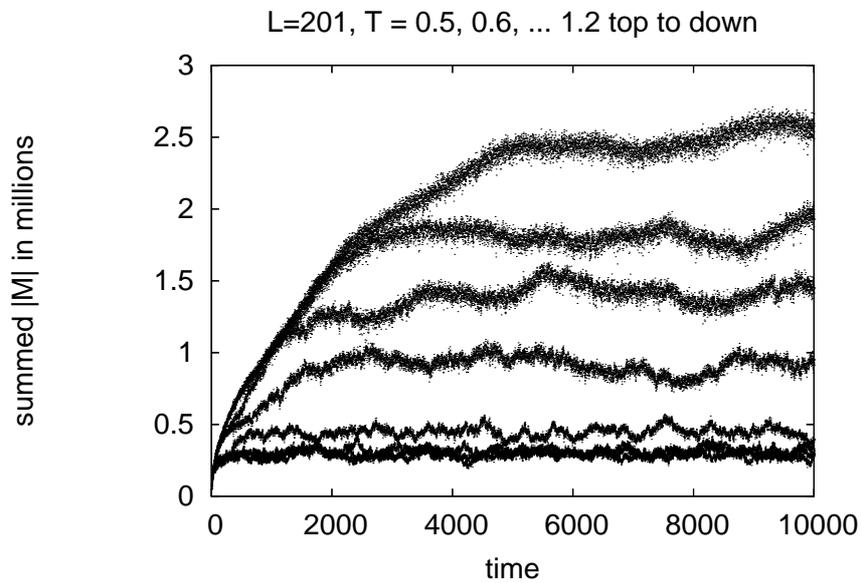}\\
\end{center}
\caption{Growth of absolute value of magnetisation if initially the spins
are up or down randomly. Average over 100 samples.
} 
\end{figure}

\begin{figure}
\begin{center}
\includegraphics[scale=0.45,angle=-90]{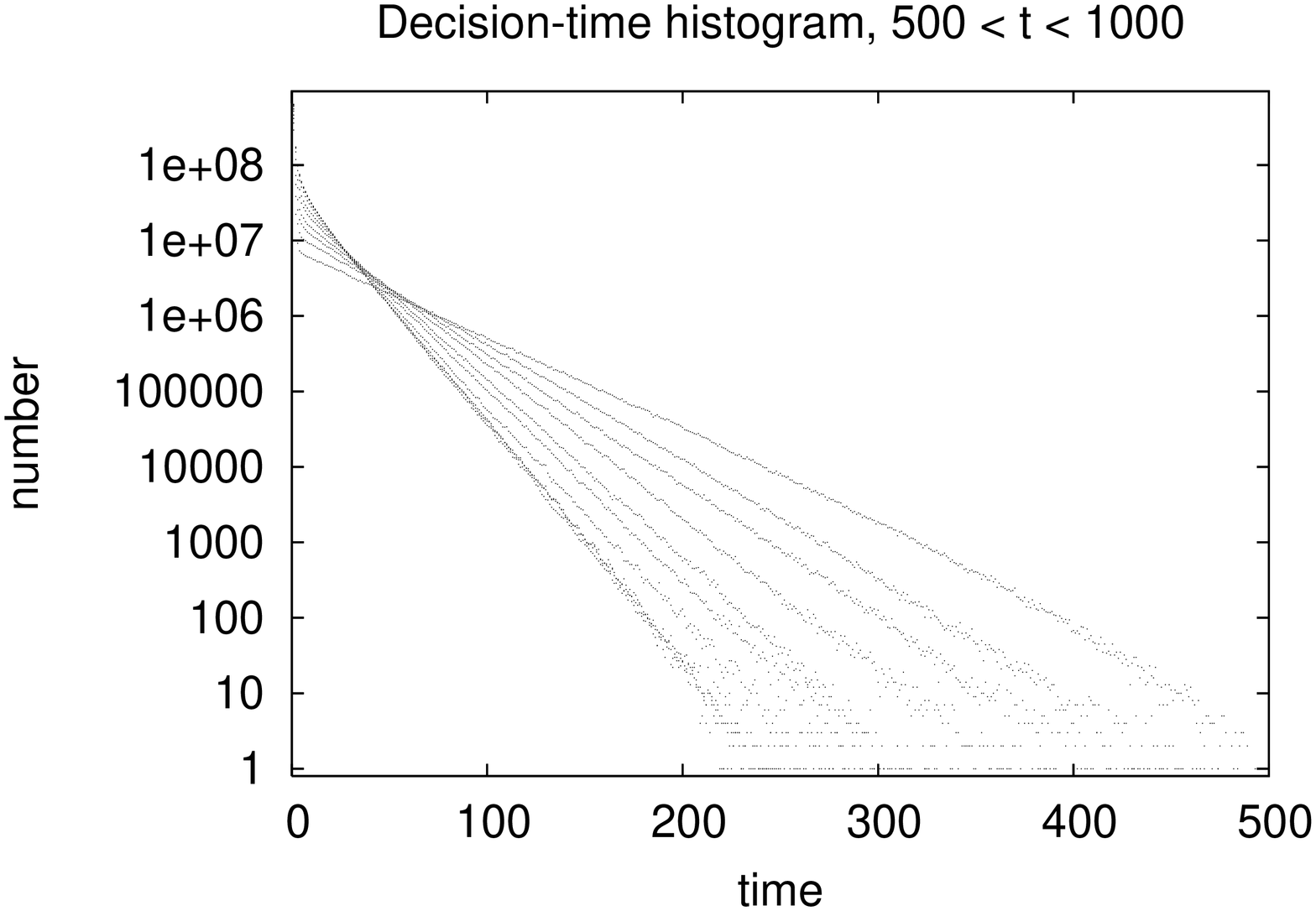}\\
\end{center}
\caption{Histograms of times between two spin flips, observed vetween 500
and 1000 iterations; $\alpha = 1, \; L = 501, \; T = 0.3, \, 0.4, \dots 1.2$.
The steeper the negative slope is, the higher is $T$. Results for $L = 1001$
and 2001 were similar (not shown). }
\end{figure}

\section{Results}

First we look at the special case $\alpha = 1$ of eq.(2b). Fig.1 shows
the summed magnetisation $\sum_i S_i$ versus time to relax exponentially
towards zero at $T=1$, when we started with all spins up. For smaller $T$ 
the magnetisation remains positive, Fig.2. Fig.3 shows its normalized
averages over the second half of the simulation.
However, these averages over many samples and/or time steps do not tell
the full story. Fig.4 shows how the system ``wants'' to recover towards
larger magnetisation but crashes back to small values if a particularly 
large noise $n$ happened. (Pictures at $T=1$ show medium-size domains of up
spins; thus if the noise flips a single spin, that spin mostly reverts 
in the next time step to its old orientation.)

The magnetisation for $T$ slightly above its critical value gets closer to
zero if we let the noise flip small plaquettes of size $b \times b$ 
instead of merely one spin. Fig.5 compares $b=1$ (previous method) with
$b$ = 3 and 5. Understandably, the critical value of $T$ is shifted downwards
if each noise event flips a whole plaquette, as also seen in Fig.5.

All figures so far referred to $\alpha = 1$; for $\alpha = 0$ and 0.5 the
results were similar, Fig.6. If for $\alpha = 1$ the noise is used with 
a probability 0.1 only, the magnetisations returned to one, up to $T = 100$
(not shown). If we start with spins randomly oriented up or down and average
over the absolute value of the magnetisation, then Fig.7 shows a spontaneous 
magnetisation to emerge after sufficiently long times for low $T$.

Finally we determined the histograms of the decision times. These are the 
time intervals between two consecutive spin flips (or opinion changes). Fig.8
shows exponential distributions: For higher $T$ the decay is faster than
for lower $T$, without evidence for critical slowing down. This kind of
decay is due to the noise.

\section{Discussion}

In a first approximation, we look at the effect of a single set of $n$ flips.
Then we observe two characteristic times of the dynamics. 
The first one is connected to the restoring of the ferromagnetic ordering. 
It is relatively short if the magnetization is far from zero; however, it can 
be arbitrarily long in the opposite case. The second time is due to the average 
lifetime of the ordered phase with a given macroscopic direction of the 
magnetisation. This time is expected to be proportional to ln$(L)$; 
sooner or later the value of the 
variable $n$ happens to be of the order of magnitude of the whole system. 
If the magnetization at that time step happens to be large enough, most of 
flipped spins were +1; therefore in this case the magnetization change is
 particularly sharp. This can be observed in an example of the time evolution of
$m$, shown in Fig.4. As for each spin the probability of flipping is $x=n/N$, 
in general case the variation of the magnetization is expected to follow 
the approximated rule for one iteration ($n$ noise events):

\begin{eqnarray}
\lefteqn{m=[n(+)-n(-)]/N \to} \nonumber\\
&&{}{[(1-x)n(+)-xn(+)-(1-x)n(-)+xn(-)]/N=}\nonumber\\
&&{}{=[(1-2x)(n(+)-n(-))]/N=(1-2x)m}\end{eqnarray}
where $n(\pm)$ is the number of spins with orientation $\pm1$. In this equation 
spins of both orientation are assumed to flip with the same probabilities. 
As we see, the state $m=0$ is the fixed point of the 
transformation. 
%However, the probability of getting there, i.e. $p(N/2)$ is
%relatively small when compared with the probability that $s>N/2$. 
%This means that a change of the sign of magnetization is more likely that its 
%reduction to zero.

More detailed inspection reveals that the two processes indicated above cannot 
be treated as independent. The exchange-mediated restoring of the ordered 
phase is slowed down even by small reductions of $m$ caused by the noise. If 
the noise intensity is small enough, some kind of dynamic equilibrium can be 
observed between the noise-induced reduction of $m$ and the increase of $m$ 
due to the exchange. Although it is difficult to speak about stationary 
processes, the observed time average in a not-too-long period of time can be 
compared to the partially ordered ferromagnetic system in temperature lower 
than its Curie temperature. In this sense, the results allow to state that the 
investigated phase transition exists also in the presence of the noise $1/n$.
As a by-product of the model we obtain the fact that any ordered phase has its 
finite lifetime. 

Now we speculate about possible social implications of these results.
Several physical concepts usually applied in sociophysics have no direct
counterparts in social systems; one of them is temperature. In physics, it 
makes sense to speak about temperature in the case of thermal equilibrium. In 
simulations, temperature measures the amount of noise; it is relevant to  
distinguish 
between the probabilities of states with different energies. In social systems 
energy is not defined and the equilibrium state is never attained. A model 
society can be considered to be rather in a self-organized critical state 
\cite{btw,bak} than in any kind of equilibrium \cite{gwb,brunk,hol,sal}.
A question arises, if results of simulations interpreted within the social sciences persist if the thermal noise is substituted by another kind of noise, 
designed as to reflect features of the self-organized criticality. In 
particular, we have found that some equivalent of the ferro-paramagnetic
phase transition persists in the presence of such a noise. 
The finite lifetime of any ordered phase can be compared to an average time 
during which a society 
supports the government formed by a given party. The $1/n$ noise in this case 
can be compared to a series of scandals which reduce the government reputation.
The size of such scandals is not limited from above; examples are at hand. 
%The aim of this paper is to answer this question.

%Other components of the system are selected as close as possible to the most 
%traditional version of a phase transition. We consider a square lattice with 
%spins $s=\pm1$ and the ferromagnetic exchange interaction $J=+1$ between each 
%spin and its four neighbours. Temperature in its usual sense is equal to zero. 
%Instead, the noise introduced here is equivalent to the flipping of $n$ 
%randomly selected spins at each time step. The probability distribution of the 
%random variable $n$ is a power law, as it is known from the theory of 
%the self-organized criticality.
%

In sociophysics, the phase transition itself is a physical trait. In
social sciences the up-down symmetry does not exist; there is always some 
bias. Then, the spontaneous symmetry breaking - the ultimate base of the model -
has again no social analogue. One could say that in sociology, the opposition is
not "plus-minus", but rather "something-nothing", related with society or its 
institutions. Then maybe we should consider the random spins equal to zero (=
non-interacting) or one rather than $\pm1$. Another physical trait is an 
interaction; in physics it joins two interacting systems symmetrically, as in 
the third Newton law, actio = -- reactio. In social systems also this symmetry 
is absent, at least on local time scale. (One may wonder to what extent the 
conquest of the Latin America had petrified feudal attitude in Spanish elites, 
making the whole country unable to develop -this would be only one example of a 
restoring of this symmetry.) These arguments indicate that further improvements 
of the model picture are needed, with a possible enrichment of the set of tools of the statistical physics.

Still, we believe that sociophysics is much too useful to be destroyed in 
this way. It should rather evolve towards a better science, more elastic in 
using physical concepts and closer to the social reality. We are going to 
reconstruct the society in mathematical models step by step. The scale-free 
$1/n$ noise, suggested by the self-organized criticality, is such a modest step 
towards this goal.

\end{document}